\documentclass[conference]{IEEEtran}
\pdfoutput=1

\usepackage[T1]{fontenc}
\usepackage[utf8]{inputenc}

\usepackage{amsmath,amssymb,amsfonts}
\usepackage[export]{adjustbox}
\usepackage{booktabs}
\usepackage{cite}
\usepackage{enumitem}
\usepackage{graphicx}
\usepackage{placeins}
\usepackage{tabularx}
\usepackage{url}
\usepackage[dvipsnames]{xcolor}

\usepackage{efbox}
\efboxsetup{linecolor=black,linewidth=0.5pt}

\usepackage{glossaries}
\glsdisablehyper
\newacronym{1ch}{\mbox{1-CH}}{single-channel}
\newacronym{1pps}{\mbox{1 PPS}}{1 pulse per second}
\newacronym{a2g}{\mbox{A2G}}{air-to-ground}
\newacronym{a2a}{\mbox{A2A}}{air-to-air}
\newacronym{cir}{\mbox{CIR}}{channel impulse responses}
\newacronym{cots}{\mbox{COTS}}{commercial off-the-shelf}
\newacronym{doa}{\mbox{DoA}}{direction-of-arrival}
\newacronym{fpga}{\mbox{FPGA}}{field programmable gate array}
\newacronym{gnssdo}{\mbox{GNSSDO}}{GNSS disciplined Oscillator}
\newacronym{ins}{\mbox{INS}}{inertial navigation system}
\newacronym{isac}{\mbox{ISAC}}{integrated sensing and communication}
\newacronym{los}{\mbox{LoS}}{line-of-sight}
\newacronym{lna}{\mbox{LNA}}{low noise amplifier}
\newacronym{mimo}{\mbox{MIMO}}{multiple-input multiple-output}
\newacronym{mqtt}{\mbox{MQTT}}{message queueing telemetry transport}
\newacronym{siso}{\mbox{SISO}}{single-input single-output}
\newacronym[plural=\mbox{SDRs},firstplural=software defined \mbox{radios (SDR)}]{sdr}{\mbox{SDR}}{software defined radio}
\newacronym{ssd}{\mbox{SSD}}{solid-state disk}
\newacronym{snr}{\mbox{SNR}}{signal-to-noise ratio}
\newacronym{rtk}{\mbox{RTK}}{real-time kinematic}
\newacronym{rpc}{\mbox{RPC}}{remote procedure call}
\newacronym{rf}{\mbox{RF}}{radio frequency}
\newacronym{rx}{\mbox{Rx}}{receive}
\newacronym{tdma}{\mbox{TDMA}}{time division multiple access}
\newacronym{tx}{\mbox{Tx}}{transmitter}
\newacronym[plural=\mbox{UAVs}]{uav}{\mbox{UAV}}{unmanned aerial vehicle}
\newacronym{ue}{\mbox{UE}}{user equipment}
\newacronym{usrp}{\mbox{USRP}}{Universal Software Radio Peripheral}
\newacronym{v2x}{\mbox{V2X}}{vehicle-to-everything} 
\newacronym{vtol}{\mbox{VTOL}}{vertical take-off and landing}
\newacronym{vrus}{\mbox{VRUs}}{vulnerable road users} 

\usepackage{hyperref}
\hypersetup{
 	pdfauthor={Julia Beuster},
 	pdftitle={Real-Time Sounding in ISAC networks: Design and Implementation of a Multi-Node Testbed with Synchronized Airborne and Ground-Based Sensors},
     pdfkeywords={Integrated sensing and communication (ISAC), 6G, distributed multi-sensor network, propagation measurements, channel sounding, unmanned aerial vehicle, air-to-air (A2A), air-to-ground (A2G), radar sensing, software defined radio (SDR), USRP, real-time kinematic (RTK)}
 }

\usepackage{orcidlink}

\title{Real-Time Sounding in ISAC networks: Design and Implementation of a Multi-Node Testbed with Synchronized Airborne and Ground-Based Sensors} 

\IEEEoverridecommandlockouts

\author{%
    \IEEEauthorblockN{%
        Julia~Beuster\raisebox{.5ex}{\orcidlink{0000-0003-1887-4278}}\IEEEauthorrefmark{1},
        Carsten~Andrich\raisebox{.5ex}{\orcidlink{0000-0002-4795-3517}}\IEEEauthorrefmark{1},
        Sebastian~Giehl\raisebox{.5ex}{\orcidlink{0009-0008-1672-1351}}\IEEEauthorrefmark{1},
        Marc~Miranda\raisebox{.5ex}{\orcidlink{0000-0000-0000-0000}}\IEEEauthorrefmark{1}, 
        Lorenz~Mohr\raisebox{.5ex}{\orcidlink{0009-0008-2599-6891}}\IEEEauthorrefmark{1}, \\
        Dieter~Novotny\raisebox{.5ex}{\IEEEauthorrefmark{2}},
        Tom~Kaufmann\raisebox{.5ex}{\IEEEauthorrefmark{3}},
        Christian~Schneider\raisebox{.5ex}{\orcidlink{0000-0003-1833-4562}}\IEEEauthorrefmark{1},
        Reiner~S.~Thomä\raisebox{.5ex}{\orcidlink{0000-0002-9254-814X}}\IEEEauthorrefmark{1}
    }
    \IEEEauthorblockA{\IEEEauthorrefmark{1}Technische Universität Ilmenau, Institute for Information Technology, Ilmenau, Germany, \\ 
    \IEEEauthorrefmark{2}AeroDCS Gmbh, Koblenz, Germany, 
    \IEEEauthorrefmark{3}CiS GmbH, Rostock, Germany
    }    
    \thanks{This research has been partially funded by the Federal Ministry of Education and Research of Germany in the projects ``6G-ICAS4Mobility'' (grant number: 16KISK241).}
}

\begin{document}
\maketitle

\begin{abstract}
As \gls{isac} capabilities become more prevalent in the mobile 6G~radio landscape, there is a substantial opportunity to enhance situational awareness across diverse applications through multi-static radar sensing within meshed ISAC networks.
To facilitate the development and testing of detection and localization algorithms across diverse scenarios, this paper introduces a synchronized distributed channel sounding testbed with airborne and ground-based multi-channel transceiver nodes with centimeter-level positioning accuracy enabled by \gls{rtk} and \gls{ins} data.
Our modular experimental measurement system is designed to include stationary sensor nodes and light-weight to medium-weight mobile nodes deployable on \glspl{uav}, cars, pedestrians, and cyclists.
Utilizing \gls{cots} hardware, specifically \mbox{\glspl{sdr}}, the testbed encourages reproducibility in academic research laboratories.
We detail the individual modules and integration steps required to achieve the specified performance.
The testbed's capabilities are validated through a real-world measurement campaign, including stationary and flying sensor nodes, aimed at detecting radar targets such as \gls{vtol} aircrafts, small hexacopters, cars and \gls{vrus} in \gls{a2a} and \gls{a2g} scenarios.
\end{abstract}


\section{Introduction}
\begin{figure}[b!]
	\includegraphics[width=\linewidth] {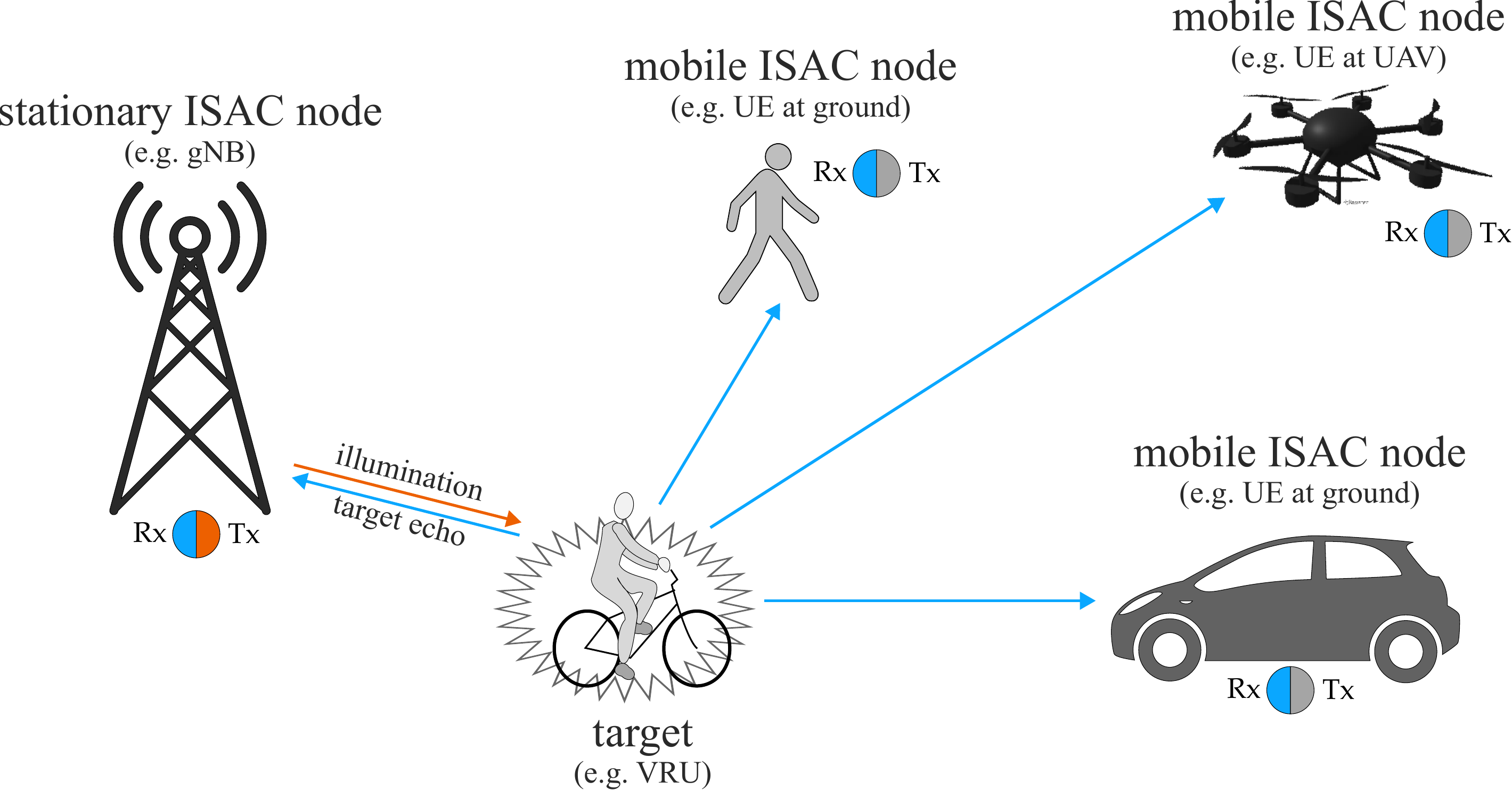}
	\caption{Simplified multi-sensor ISAC network with stationary and mobile airborne and ground-based transceiver nodes, where the Tx node illuminates a target to allow radar sensing at the Rx nodes. Typically, all participating nodes can operate as Rx and Tx.  
    }
\label{isacnetwork}
\end{figure}

The upcoming integration of \gls{isac} capabilities into a broad range of mobile radio devices (6G and beyond) offers a promising pathway to enhance situational awareness and communication efficiency across various applications.
Leveraging the advantages of multi-static sensing within meshed ISAC networks, as introduced in~\cite{10289611}, these multi-sensor \gls{isac} architectures comprise various transceiver nodes that act as illuminators and sensors within a distributed infrastructure to allow detection and tracking of position-related parameters of static and moving objects, as illustrated in  Fig.~\ref{isacnetwork}.
A crucial element of this envisioned integration is the development of channel sounding~\cite{843078} testbeds that include all type of nodes present in such an multi-sensor \gls{isac} network, facilitating the testing of \gls{isac} algorithms across various communication scenarios.
A significant hurdle lies in the design and implementation of the mobile airborne sensor nodes to capture the rapid variations and unique propagation characteristics encountered in \gls{uav} flight scenarios, especially given the stringent weight and space constraints present at \gls{uav}, as comprehensively discussed in the survey by Mao~et~al.~\cite{10616106}.
Ideally, such an experimental testbed should be constructed using \gls{cots} hardware to ensure availability across different academic research laboratories and enable customization to various measurement scenarios.
\begin{table*}[h!]
	\caption{Comparison of SDR-based channel sounding measurement systems with UAV nodes}
	\centering
    \addtolength{\tabcolsep}{-0.25em}
    \begin{tabularx}{\linewidth}{Xllllll}
	\toprule
    \small \textbf{Author} & \textbf{Zhang} \cite{9135251} & \textbf{Choi} \cite{9771649} & \textbf{Beuster} \cite{10133118} & \textbf{Zhang} \cite{zhang2023channel} & \textbf{This work} \par\vspace{0.2em}\\
    Year & 2020 & 2022 & 2023 & 2023 & 2025\\
    \midrule
    Mobile (airborne) node & Tx-only & Tx-only & Tx-only & Tx-only & Tx/Rx & Tx/Rx \par\vspace{0.6em}\\
    \quad -- SDR-based & 1$\times$USRP N210 & 1$\times$USRP B210 & 1$\times$USRP B205 & 1$\times$USRP X310 & 1$\times$USRP X310, & 1$\times$USRP B205 \\
     \quad \quad RF hardware & & & & & 1$\times$USRP B205 \par\vspace{0.2em}\\
    \quad -- number of channels,& 1-CH 25 MHz & 1-CH 56 MHz & 1-CH 40 MHz\footnotemark[1] & 1-CH 80 MHz\footnotemark[1] & 4-CH 100 MHz, & 1-CH 56 MHz\\
    \quad \quad bandwidth & & & & & 1-CH 56 MHz \par\vspace{0.2em}\\
    \quad -- load capacity & unknown & < 5.5\,kg & < 800\,g & < 8\,kg & < 5\,kg & < 2\,kg \par\vspace{0.2em}\\
    \quad -- power supply & battery-powered & battery-powered & battery-powered & battery-powered & mains-/ battery- & battery-powered\\
    \quad \quad \scriptsize(battery life) & \scriptsize (unknown) & \scriptsize(15\,min) & \scriptsize(5\,hours) & \scriptsize (unknown) & powered\footnotemark[2]\scriptsize(45\,min) & \scriptsize (4\,hours) \par\vspace{0.2em}\\
    \quad -- synchronization & GPSDO \scriptsize(TCXO) & GPSDO \scriptsize(TCXO) & no sync & no sync & GPSDO \scriptsize(OCXO), & GPSDO \scriptsize(OCXO),\\
    & & & & & REF IN switch\footnotemark[3], & REF IN switch\footnotemark[3] \par\vspace{0.2em}\\
    \quad -- positioning & GPS & unknown & RTK GNSS & unknown & RTK GNSS, INS & RTK GNSS, INS\\
    \midrule
    Stationary node & Rx-only & Rx-only & Tx/Rx & Rx-only & Tx/Rx \par\vspace{0.6em}\\
    \quad -- RF hardware & 1$\times$USRP N210 & proprietary & N$\times$USRP X310 & 1$\times$USRP X310 & N$\times$USRP X310 \par\vspace{0.2em}\\
     & 1-CH 25 MHz & & 4-CH 100 MHz & 1-CH 100 MHz\footnotemark[1] & 4-CH 100 MHz \par\vspace{0.2em}\\
    \quad -- synchronization & GPSDO \scriptsize(TCXO) &  & GNSSDO \scriptsize(Rb) & no sync & GNSSDO \scriptsize(Rb)\par\vspace{0.2em} \\
    \quad -- positioning & GPS & unknown & RTK GNSS & unknown & RTK GNSS \\
    \midrule
    Measurement scenario & A2G & A2G & A2G, V2X & A2G & \multicolumn{2}{l}{A2A, A2G, V2X} \\
    & & \scriptsize200$\times$400m & \scriptsize500$\times$500m & & \scriptsize100$\times$100m\par\vspace{0.2em}\\ 
    
    \quad -- signal parameters & pseudo-noise & OFDM & OFDM & pseudo-noise & OFDM \\
     & \scriptsize 20\,MHz $@$2.585\,GHz & \scriptsize 46\,MHz $@$3.5\,GHz & \scriptsize 32\,MHz $@$3.75\,GHz & \scriptsize 80\,MHz $@$3.6\,GHz & \multicolumn{2}{l}{\scriptsize 80\,MHz/48\,MHz $@$3.75\,GHz} \par\vspace{0.2em}\\
    \quad -- calibration & B2B & & B2B & B2B & B2B\par\vspace{0.2em}\\
    \quad -- number of nodes & 1$\times$Tx, 1$\times$Rx & 1$\times$Tx, 1$\times$Rx & 1$\times$Tx, 3$\times$Rx & 1$\times$Tx, 1$\times$Rx & \multicolumn{2}{l}{2$\times$Tx, 7$\times$Rx, 1$\times$Tx/Rx} \par\vspace{0.2em}\\
    \quad -- radar targets\footnotemark[4] & & & UAV \scriptsize(hexacopter) & & \multicolumn{2}{l}{car, cyclist, UAV \scriptsize(VTOL, quadcopter)}\\
    \bottomrule
	\multicolumn{6}{l}{\footnotemark[1] \scriptsize Bandwidth limited by Host PC's streaming capabilities.}\\
    \multicolumn{6}{l}{\footnotemark[2] \scriptsize Live-swap design to allow power supply replacement without interruption of the operating measurement system, for instance to consider warm-up times.}\\
    \multicolumn{6}{l}{\footnotemark[3] \scriptsize Circuit design to allow switching between 10 MHz and 1 PPS reference signals for single REF IN port at USRP B205mini-i.}\\
    \multicolumn{6}{l}{\footnotemark[4] \scriptsize With reliable ground truth data by RTK GNSS modules.}\\
 	\end{tabularx}
 \label{tab:soa}
 \end{table*}
Recent studies~\cite{9135251, 9771649, 10133118, zhang2023channel} introduced the implementation of channel sounding systems tailored for \gls{uav} applications using \gls{cots} hardware in form of \glspl{sdr}.
However, these proposed systems are limited to \gls{1ch} transmit-only~(Tx-only) nodes, as implementing \gls{rx} paths presents significant challenges in hardware design, particularly regarding recording reliability. 
Consequently, this limitation restricts research to \gls{a2g} scenarios, excluding \gls{a2a} scenarios essential for \gls{isac} applications that include multiple \glspl{uav}, for instance in drone swarms.
To overcome these limitations, our research aims to advance the proposed channel sounding techniques by extending the \gls{isac} testbed with synchronized, multi-channel airborne transceiver (Tx/Rx) nodes.
Table~\ref{tab:soa} provides a comparative analysis of recent work, highlighting their strengths and weaknesses in contrast to our proposed system. 
With this paper, we introduce a synchronized distributed channel sounding testbed, designed to investigate and develop algorithms for multi-sensor \gls{isac} networks, featuring both airborne and ground-based nodes.
The proposed measurement system is modular and includes SDR-based stationary and mobile transceiver nodes, ranging from light-weight to medium-weight, with centimeter-level \gls{rtk} and \gls{ins}-based positioning accuracy.
We will detail the individual testbed modules and essential integration steps required to achieve the specified performance and accuracies. 
The testbed's performance is validated through an exemplary real-world measurement campaign, including stationary sensor nodes on a rooftop and on the ground, as well as mobile sensor nodes mounted on cars and transport hexacopters.
This campaign aims to detect radar targets such as \gls{vtol} aircrafts, small hexacopters, cars, and \gls{vrus} in form of cyclists, and can be used to generate publicly accessible measurement datasets for the development and evaluation of future \gls{isac} algorithms.
The focus of this publication is on the measurement system itself, rather than the measurement data sets and results.

\section{Multi-node sounding testbed}

\begin{figure*}[ht]
	\includegraphics[width=\linewidth] {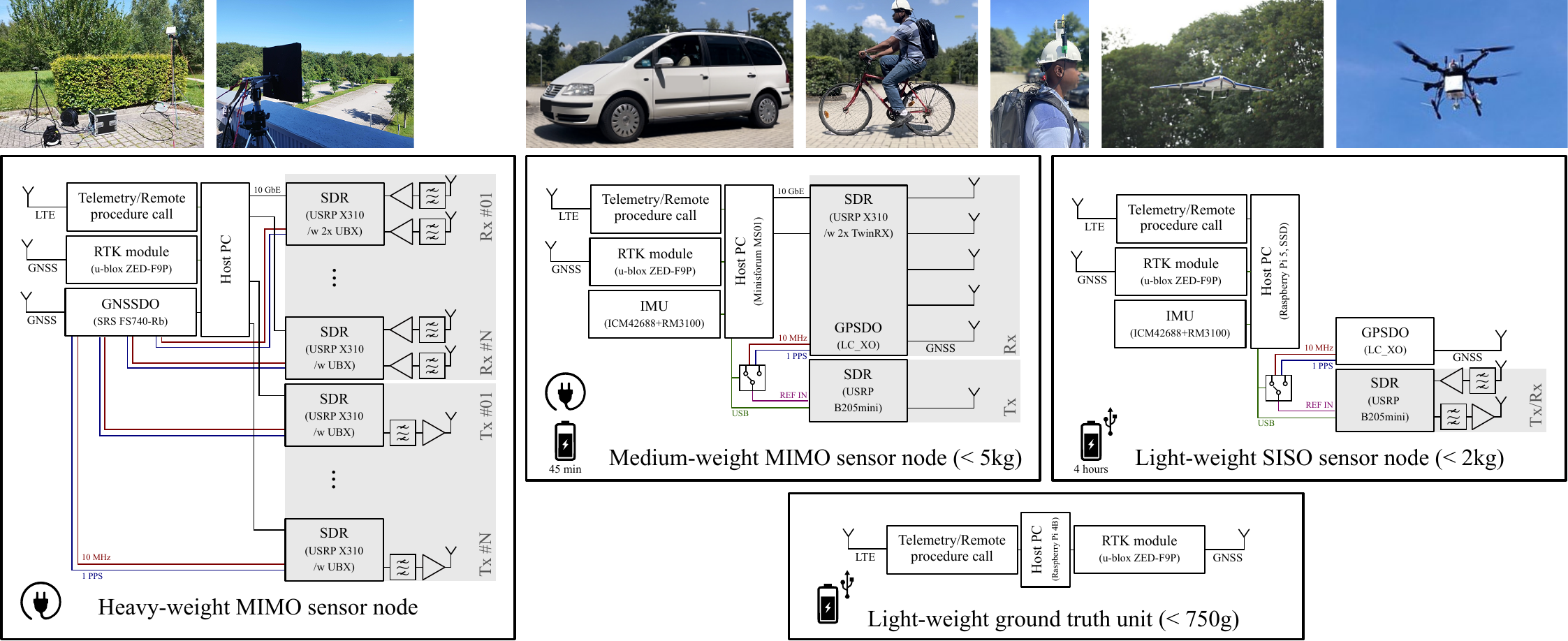}
	\caption{Simplified block diagrams of the various node types in a COTS-based channel sounding testbed for multi-sensor ISAC networks. The multi-node architecture is modular with a variable number of distributed synchronized transceiver nodes, each featuring configurable signal parameters and switchable receive and transmit paths, as well as a highly accurate ground truth in terms of positioning for all participating nodes.
    }
\label{setup}
\end{figure*}

Considering the multitude of future \gls{isac} algorithms, our testbed prioritizes highly reliable, continuous data recording of \gls{cir} with large bandwidths. 
This approach supports applying subsequent signal processing for comparing and optimizing algorithms, rather than focusing on real-time \gls{isac} functionality, thereby avoiding the cost and operational effort of repetitive measurement campaigns.
The measurement system needs to be modular with a variable number of synchronized transceiver nodes, featuring configurable signal parameters and switchable receive and transmit paths, and an highly accurate ground truth in terms of positioning and orientation for all participating nodes.
All components are constructed from \gls{cots} hardware to allow easy scaling based on experimental needs.
Fig.~\ref{setup} illustrates the key components of the measurement platform.
The system comprises 
\begin{enumerate}
    \item heavy-weight multi-channel sensor nodes, that can be used as stationary nodes or installed at large vehicles~\cite{10133118},
    \item medium-weight (less than 5\,kg) multi-channel sensor nodes, suitable for mobile use at cars, \gls{vrus}, and \glspl{uav}, 
    \item small, light-weight (less than 2\,kg) single-channel sensor nodes, also for mobile use on the ground and in the air, 
    \item small, light-weight (less than 750\,g) ground truth units for ground-based and airborne objects, such as radar targets. 
\end{enumerate}
All nodes are designed for easy transport, deployment, operation in almost arbitrary positions using wireless telemetry and \gls{rpc}.
Each mobile node is equipped with a self-sufficient power supply capable of operating for at least 45~minutes, matching typical flight times, with a live-swap option to replace the power supply without interrupting system operation, accommodating hardware warm-up times.
The mobile nodes also feature on-board data storage to eliminate the need for mechanical attachments such as cables from RF hardware in the air to a data storage unit on the ground.

\subsection{RF Hardware}
The testbed is built around \gls{rf} hardware in form of \glspl{sdr} from the \gls{usrp} device family, chosen for their high degree of flexibility and popularity within academic research.
Two specific devices from the product family were utilized in this paper: 
the half-width~1U rack-mountable \gls{mimo} USRP~X310, which features two slots for RF frontends and supports streaming of up to 160~MHz of baseband bandwidth to/from host PCs using dual 10~GbE, 
and the small, light-weight \gls{siso} USRP~B205mini-i, which enables streaming of up to 56~MHz from/to host PCs using USB~3.0.
Both devices feature individual receive and transmit paths that can be activated and deactivated using the cross-platform UHD driver.

\subsubsection{Heavy-weight MIMO sensor node}
This node utilizes USRP~X310 equipped with UBX daughterboards as dual-channel transceiver platforms, each connected via 10~GbE to a high-performance host server configured with \glspl{ssd} for data recording, as introduced in~\cite{10133118}.
To optimize the signal quality and minimize harmonics and spurious emissions, each transmit path includes a band-pass filter and a power amplifier.
On the receiver side, band-pass filters are deployed in conjunction with a \gls{lna} to minimize the noise figure and mitigate the influence of adjacent interferers.
The number and configuration of these nodes are scalable and limited by available hardware resources.

\subsubsection{Medium-weight MIMO sensor node}
This sensor node utilizes one USRP~X310 with two TwinRX~frontends, offering a 4-channel receiver with phase synchronization, which is for instance essential for direction-finding applications, and a high dynamic range, eliminating the need for an additional \gls{lna}~\cite{9769488}.
Alternatively, other \gls{rf} frontends such as the UBX~daughterboards can be employed in the USRP~X310 to support transmit and receive operation if required.
This receiver node further can be extended by an additional USRP B205mini-i transmitter to provide a reference signal essential for enhancing the accuracy and reliability of \gls{doa} estimation.
The USRP~X310 is connected via 10~GbE Ethernet, and the USRP~B205mini-i via USB~3.0, to a compact high-performance host PC, specifically the Minisforum MS01. 
This PC weights only 850~g, while featuring dual 10~GbE ports and NVMe~SSDs for baseband signal streaming and storage.

\subsubsection{Light-weight SISO sensor node}
Designed for applications requiring minimal weight and power consumption, this node incorporates the USRP~B205mini-i connected via USB~3.0 to a Raspberry~Pi~5 serving as host PC.
In comparison with the Tx-only node introduced in~\cite{10133118}, the Raspberry~Pi~5 includes a faster CPU and an NVMe~SSD for baseband signal streaming and storage, while maintaining a small form factor and a weight of only 125~g.
The transmit and receive path each include a band-pass filter and a compact, USB-powered~\gls{lna}.

\subsection{Distributed Synchronization}
In channel sounding testbeds, synchronization is a critical aspect that ensures the accuracy and reliability of the measurement data and can be broadly divided into two categories: real-time synchronization, that is for instance crucial for switching transceiver nodes in \gls{isac} networks, and post-processing synchronization accuracy, which involves drift compensation in recorded measurement data.

\subsubsection{Real-time synchronization}
Real-time synchronization is essential for various measurement applications, that require precise timing to coordinate the transmission and reception of signals, for instance for Tx/Rx switching in \gls{isac} networks and using \gls{tdma} schemes for multi-static radar sensing.
\glspl{gnssdo} typically provide \gls{1pps} and 10~MHz reference signals synchronizing the clocks of all RF hardware components within the distributed nodes and enabling straightforward synchronization at almost any location with a clear sky view.
As introduced in our previous work~\cite{10133118}, the real-time synchronization of the heavy-weight sensor nodes can be easily achieved using high-precision timing references such as the Stanford Research Systems~FS740 with internal rubidium time base.
Implementing synchronization for the mobile nodes, on the other hand, faces additional challenges due to their use in dynamic scenarios, especially when deployed on airborne objects, as well as limitations imposed by slimmed-down RF hardware.
To meet the weight and space constraints, a compact, light-weigth \gls{gnssdo}, previously characterized for \gls{uav} deployment~\cite{10722561}, the Jackson Labs GPSDO~LC$\_$XO~OCXO, is chosen as a compromise between a small form-factor and comparatively moderate reference signal stability and due its availability to the authors.
This \gls{gnssdo} provides both \gls{1pps} and 10~MHz signals to the SDRs in the light-weight and medium-weight sensor node.
However, the small \gls{sdr} \mbox{USRP B205mini-i} has only one reference input~(REF~IN) port, which can accept either the 10~MHz or 1~PPS reference signal.
To address this, a remote-controlled RF switch circuit with a latched relay was designed.
This circuit, which can be controlled via the Raspberry~ Pi~5 GPIO header or USB, allows software-driven toggling between the 1 PPS and 10 MHz reference signal ensuring both initial absolute time synchronization and ongoing relative time synchronization during the measurement.
Minor modifications to the original \gls{fpga} image of the USRP~B205mini-i were made to prevent resettling after switching the reference signal source.
The circuit design and \gls{fpga} image modifications required are available from our GitHub account~\cite{github}.

\begin{figure}[t]
\centering
	\includegraphics[width=\linewidth] {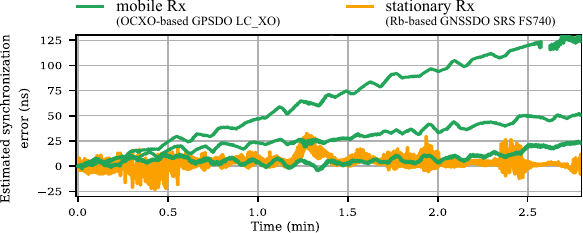}
	\caption{Time-variant synchronization error illustrated by the LoS delay estimation between one Tx (light-weight sensor node mounted at a hovering UAV) and all Rx nodes (stationary heavy-weight sensor nodes and mobile light-weight sensor nodes at hovering UAVs), as shown in scenario Fig.~\ref{measurement_site}, matching the expected performance range of the used GNSSDOs~\cite{10133118, 10722561}. High-resolution parameter estimation was applied to estimate the strongest path in the multi-path-affected Rx signals.}
\label{sync_los}
\end{figure}

\subsubsection{Synchronization accuracy after post-processing}
While real-time synchronization is crucial for immediate operations, post-processing synchronization accuracy is essential for using the recorded sounding data. 
Over time, the reference signal standards experience oscillator drift due to environmental factors, such as temperature changes, mechanical disturbances and GNSS signal impairments, that can differ significantly between individual sensor nodes, as shown in Fig.~\ref{sync_los}.
This drift can introduce errors into the recorded data in form of sampling clock and carrier frequency offsets, both affecting the accuracy of subsequent analyses and localization algorithms.
To compensate for this time-variant drifts, typically post-processing techniques are employed by utilizing a temporary beacon Tx with unobstructed LoS to all RXs, as shown in~\cite{10133118}.
These techniques involve estimating the synchronization error in the recorded data using high-resolution parameter estimation frameworks to estimate the propagation delay of the \gls{los} component and comparing the estimated delay with the expected delay from the ground truth data.
This approach mitigates phase and timing errors induced by the time-variant carrier frequency and sampling clock offsets, so that the post-processed sounding data models the actual physical channel with a greater degree of accuracy.

\subsection{Positioning and Orientation}
In measurement networks that incorporate flying nodes, the integration of orientation information in conjunction with  precise positioning data is crucial for the accuracy of localization outcomes. 
The 6~degrees of freedom~(DOF), while maneuvering freely in three-dimensional space, can significantly alter antenna positions. 
Highly-precise positioning data with centimeter-level accuracy can be obtained from low-cost \gls{rtk} GNSS receivers like the u-blox ZED-F9P, using real-time radio technical commission for maritime services~(RTCM) correction data from a SAPOS® system, as shown in~\cite{10133118}, to provide the necessary spatial resolution.
Orientation information, derived from an common commercial UAV navigation \gls{ins} module combining the ICM42688 gyroscope and accelerometer with the RM3100 magnetometer, complements the positioning data by providing the roll, pitch, and yaw angles of the dynamic nodes.
This combined information ensures that the exact antenna position and orientation are known at all times, even during complex flight maneuvers.

\subsection{Power supply and Housing}

\begin{figure}[t!]
	\includegraphics[width=\linewidth] {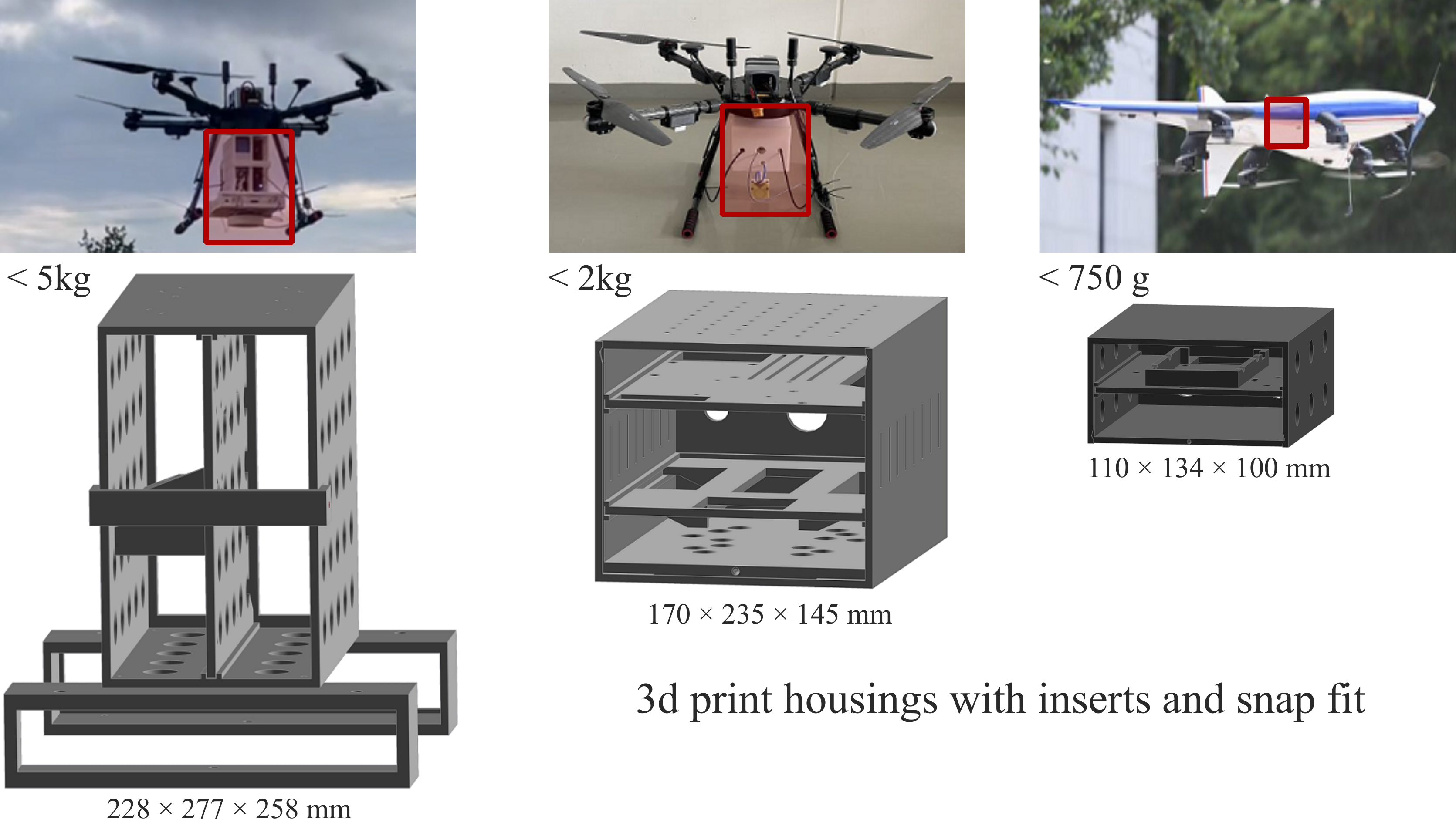}
	\caption{3d printed protective housing to mount the sensor node components to UAVs, exemplarily a small transport hexacopter and a VTOL.}
\label{3dprints}
\end{figure}

\subsubsection{Stationary nodes}
Each sensor node's hardware consists of mains-powered appliances that can be mounted in 19''\,racks for easy transport and deployment.

\subsubsection{Mobile nodes}
Ensuring that the mobile sensor nodes minimize the overall weight and space while maintaining their functionality and robustness is crucial for their effective deployment in diverse dynamic measurement scenarios.
The medium-weight and light-weight sensor nodes are designed as stand-alone systems, eliminating the need for mechanical attachments to stationary objects, while considering physical protection, thermal management of sensitive hardware, and quick installation and removal for swift deployment on different carriers, such as in backpacks of \gls{vrus} as pedestrians and cyclists, at small cars or on common transport \glspl{uav}.

\paragraph{Power supply}
To ensure a consistent power supply, reduce the risk of malfunctions in the sensitive measurement hardware due to voltage fluctuations and electromagnetic interferences, and avoid time-consuming battery changes with repetitive system warm-up periods in between flights, all components of the mobile sensor node are powered by a self-sufficient power supply instead of relying for instance on the UAV's battery.
The lightweight sensor node and ground truth unit are battery-powered using 400\,g or 230\,g exchangeable 5\,V~USB power banks allowing operation times of up to 4~hours.
The medium-weight sensor node features a live-swap power supply unit for battery replacement and changing between battery-powered and mains-powered operation.
This enables for instance hardware warm-up periods using mains-powered 230\,V and starting a measurement flight by switching to the 815\,g exchangeable 24\,V LiPo battery with operation times of up to 45\,minutes.
The circuit board design is available at our GitHub account~\cite{github}. 

\paragraph{3d printed protective housing}
All components of the medium- and light-weight sensor node and the ground truth unit are installed in customized protective housing ensuring that the hardware is well-protected, easy to deploy, and adaptable to a wide range of measurement scenarios, as shown in Fig.~\ref{3dprints}.
Constructed using 3D printing technology, the housing incorporates light-weight materials and strategic fill-ins to minimize weight while maintaining structural integrity.
Large ventilation slots are positioned at non-load-bearing areas to enhance thermal management and reduce overall weight.
The design features the use of inserts instead of fixed mounts enabling quick and easy installation and removal on various carrier systems.
To further enhance usability, the housing employs a tool-less operation design with snap-fit closures, which simplifies the assembly process and reduces wear and tear on sensitive measurement components, as they are mounted securely at the inserts and do not require frequent handling.
The 3d print design files and additional mounting instructions are available on our GitHub account~\cite{github}, ensuring reproducibility and further customization by other researchers.

\subsection{Measurement Software}
The measurement software is designed to record data for offline signal processing, rather than real-time localization, aligning with the testbed's focus on enabling detailed analysis and algorithm development.
The reliable USRP software is responsible for synchronized operation of all \glspl{sdr} with a \gls{rpc} interface that allows operators to remotely mute/unmute the Tx paths and start/stop the recording of complex baseband signal at the Rx, as introduced in~\cite{10133118}.
This is enhanced using real-time telemetry for continuous supervision and seamless operation.
Separate, reusable Python scripts are used to handle telemetry for the numerous \glspl{gnssdo}, \gls{rtk} GNSS receivers, and parameters such as CPU and SSD usage and system temperature for all participating nodes.
These scripts run as auto-started system services, ensuring that telemetry data is continuously collected and transmitted via the \gls{mqtt} protocol.
A central server processes the telemetry data, which is sent in form of a single Rx signal period and software telemetry at an update rate of 1 Hz, into real-time spectrum and impulse response plots, providing operators with visual feedback for RF function supervision.
Additionally, an interactive map for GNSS function supervision and a telemetry table for color-coded monitoring of over 200 system parameters are generated.
All this is accessible via responsive web applications, allowing operators to monitor the system from mobile devices through a web browser.
Given the system's complexity and the critical nature of the costly and time-consuming large-scale measurement campaigns with a high amount of moving nodes, this real-time telemetry is essential to ensure continuous monitoring and prevent undetected failures from compromising the results.

\section{Real-world measurement campaign}

\begin{table}[t!]
	\caption{Summary of measurement campaign characteristics.}
	\vspace{-1em}
	\centering
    
    \begin{tabularx}{\linewidth}{Xl}
	\toprule
    \small Measurement site & \\
    \quad -- 2-story building at parking lot & $\sim$100$\times$100m\\
	\quad -- RTK base station & SAPOS\textsuperscript{\textregistered}\cite{sapos}\\

    \midrule
 	\small Signal parameters & \\
 	\quad -- OFDM & Newman phase sequence \cite{6053}\\
 	\quad -- Center frequency & 3.75\,GHz \\
    \quad -- Symbol length & 16 \textmu s \\
 	\quad -- Bandwidth (used) & 80 MHz, 48 MHz \\
    \quad -- Subcarriers & 1280, 768 \\
	\quad -- Effective isotropic radiated power & 46 dBm, 23 dBm \\
    \quad -- calibration & B2B\\
    \midrule
    \multicolumn{2}{l}{\small Stationary nodes}\\
    \quad -- Tx\footnotemark[1] \mbox{({\color{BrickRed} $\bullet$})} & 1 (location \#01 at rooftop) \\
    \quad -- Rx\footnotemark[1] \mbox{({\color{BurntOrange} $\bullet$})} & 2 (location \#01 at rooftop) \\
    \quad  & 2 (location \#02 at ground) \\
    \midrule
    \multicolumn{2}{l}{\small Mobile nodes}\\
    \quad -- Tx-only\footnotemark[2] \mbox{({\color{MidnightBlue} $\bullet$})} & 1 (at car, cyclist or hexacopter)\\
    \quad -- Rx-only\footnotemark[2]\mbox{({\color{Green} $\bullet$})}& 3 (at car or hexacopter)\\
    \quad -- Tx and 4-CH Rx\footnotemark[3] \mbox{({\color{LimeGreen} $\bullet$})} & 1 (car-mount)\\
    \midrule
    \multicolumn{2}{l}{\small Targets\footnotemark[4]}\\
    \quad -- at ground level & mid-sized car, cyclist \\
    \quad -- airborne & VTOL, quadcopter \scriptsize(DJI Matrice) \\
    \bottomrule
	\multicolumn{2}{l}{\footnotemark[1] \scriptsize Heavy-weight MIMO sensor node with tiltable, directional antennas with }\\
    \multicolumn{2}{l}{\quad \scriptsize10~dB beamwidth at 40 degree.}\\
    \multicolumn{2}{l}{\footnotemark[2] \scriptsize Light-weight SISO sensor node with omni-directional antenna.}\\
    \multicolumn{2}{l}{\footnotemark[3] \scriptsize Medium-weight MIMO sensor node with antenna array.}\\
    \multicolumn{2}{l}{\footnotemark[4] \scriptsize With light-weight ground truth unit.}\\
 	\end{tabularx}
 \label{tab:specs_meas}
 \end{table}

 \begin{figure}[t!]
	\includegraphics[width=\linewidth] {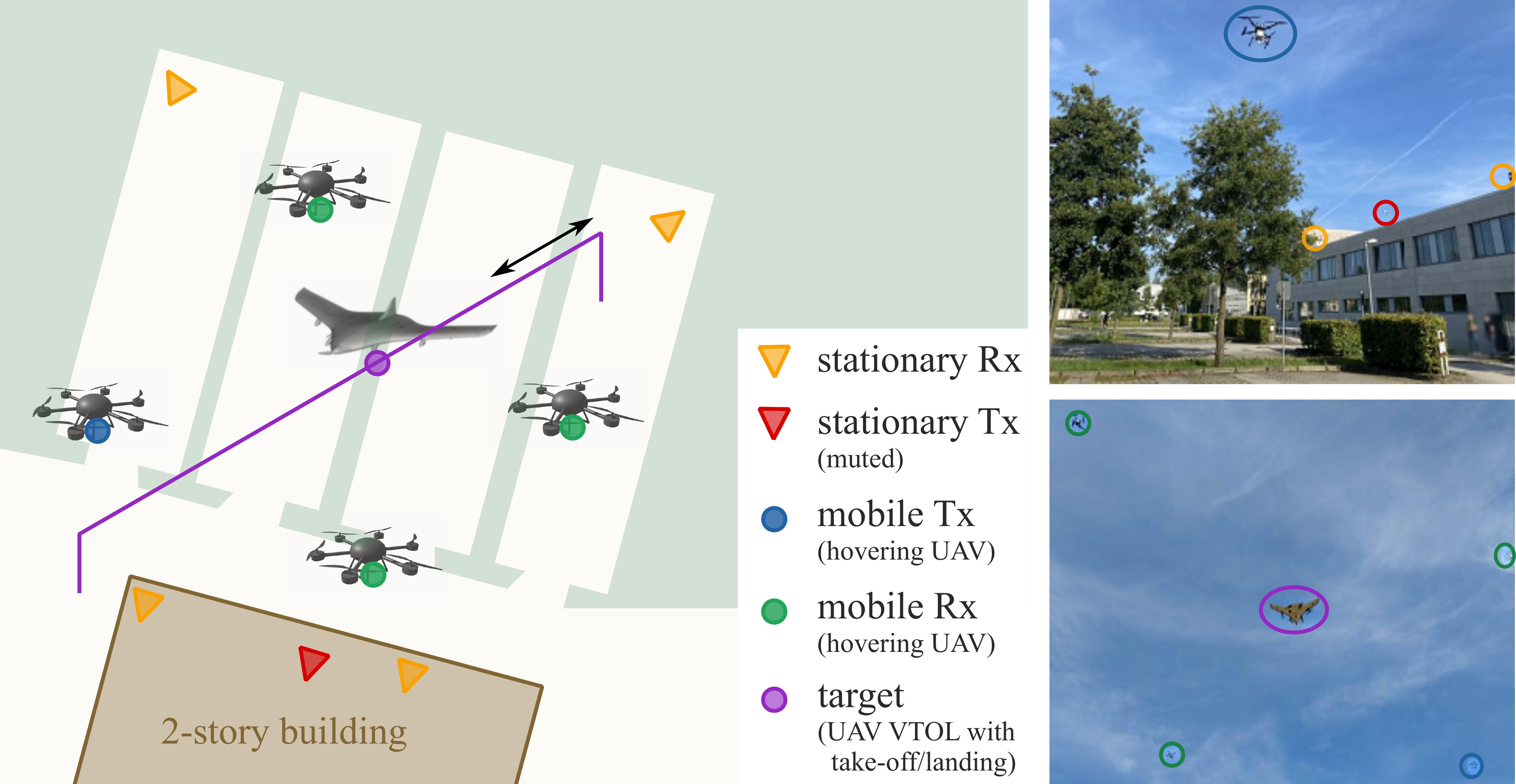}
	\caption{Simplified illustration of one exemplary scenario of the real-world measurement campaign with stationary sensor nodes \mbox{(Rx: {\color{BurntOrange} $\bullet$}, Tx: {\color{BrickRed} $\bullet$})} and mobile sensor nodes at hovering \glspl{uav} \mbox{(Rx: {\color{ForestGreen} $\bullet$}, Tx: {\color{RoyalBlue} $\bullet$})} to detect a VTOL aircraft \mbox{(target: {\color{Purple}$\bullet$})} in arrival and departure situation in terms of radar sensing.
    }
\label{measurement_site}
\end{figure}

To demonstrate the capabilities of our experimental testbed for multi-static radar sensing in \gls{isac} networks, we conducted a field test to exemplarily show channel sounding data set acquisition in diverse multi-node V2X, \gls{a2a}, and \gls{a2g} measurement scenarios for different ground-based and airborne targets, covering an area of approximately 100$\times$100~m at a \mbox{2-story} building. 
The sensor node setup, antenna configuration, trajectory planning, and signal parameters for the real-world measurement campaign were chosen to provide conditions and signal parameters similar to those present for 5G user equipment (UE) in FR1, and are summarized in Table~\ref{tab:specs_meas}.
The measurement campaign was designed to capture a diverse range of scenarios, with a focus on generating a publicly accessible dataset \cite{beuster2025datasetICAS4M} for \gls{isac} algorithm development and validation.

\textit{Mobile and infrastructure-based sensing of flying objects:}
Fig.~\ref{measurement_site} shows only a small part of the extensive measurement campaign, providing just a sneak peak of the upcoming data set. 
This specific scenario included two stationary nodes for infrastructure-based sensing: one at the rooftop of the building with one transmitter and two receivers, as well as one at the ground of the parking lot with two receivers, each equipped with tiltable, directional antennas at tripods for exposed positioning and additional absorbers to minimize LoS power and therefore, to improve receiver sensitivity and linearity.
Additionally, the setup included four mobile sensor nodes: one transmitter and three receivers using light-weight sensor nodes mounted on hovering transport hexacopters with omni-directional antennas.
The target was a flying \gls{vtol} aircraft in arrival and departure situations, as recently used in a study about of bi-static Micro-Doppler analysis in \gls{isac} frameworks~\cite{10979139}.
All transceiver constellations were calibrated using back-to-back~(B2B) calibration.
\begin{figure}[t!]
\includegraphics[width=\linewidth]{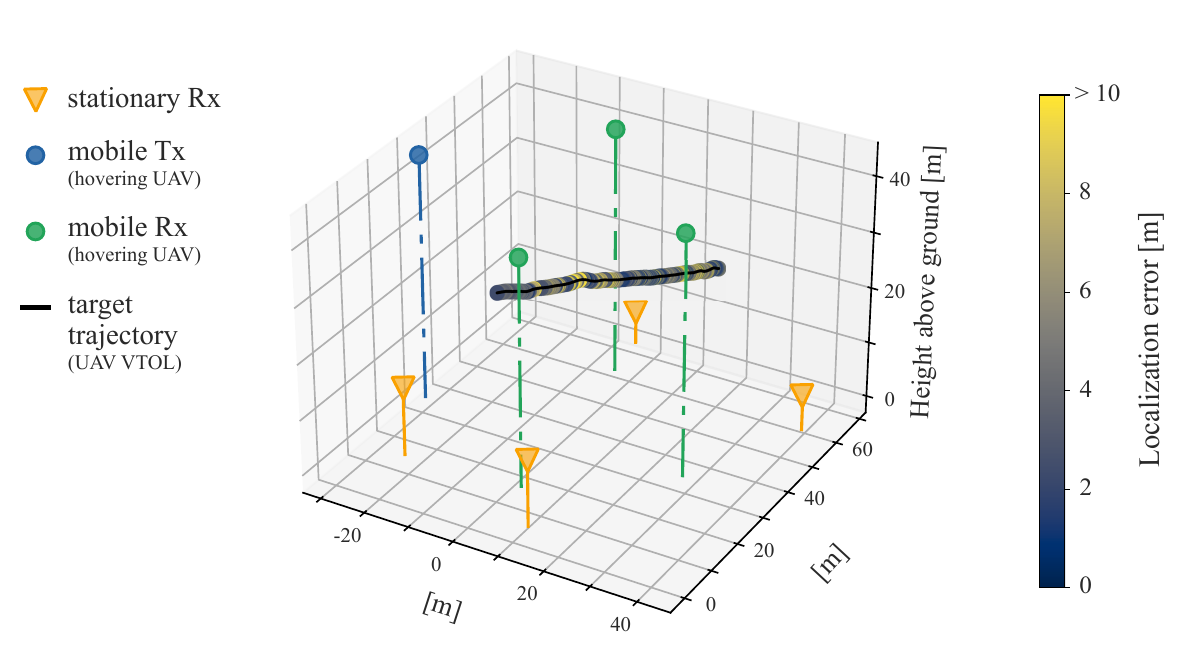}
\caption{Localization error between the estimated target positions, derived using least-squares with estimated target delays of all receiver nodes and sensor node position data, and the actual target position obtained from the ground truth unit.}
\label{measurement_localization}
\end{figure}
Given the wide assortment of radar-based localization methods, the evaluation of particular algorithms using the measurement data is beyond the scope of this paper.
The system's capability to detect a large \gls{uav} using its reflectivity is demonstrated using a simple \gls{los} situation in combination with a common delay-based radar algorithm.
To estimate the \gls{vtol} positions, the delay parameters of the target reflection are estimated at all receiver nodes and fused with the position information of the participating sensor nodes to calculate the positions using least-squares.
As shown in Fig.~\ref{measurement_site}, the \gls{vtol} is detected on its flight route mainly limited by the antenna constellation and surrounding vegetation. 
\begin{figure}[t!]
\includegraphics[width=\linewidth]{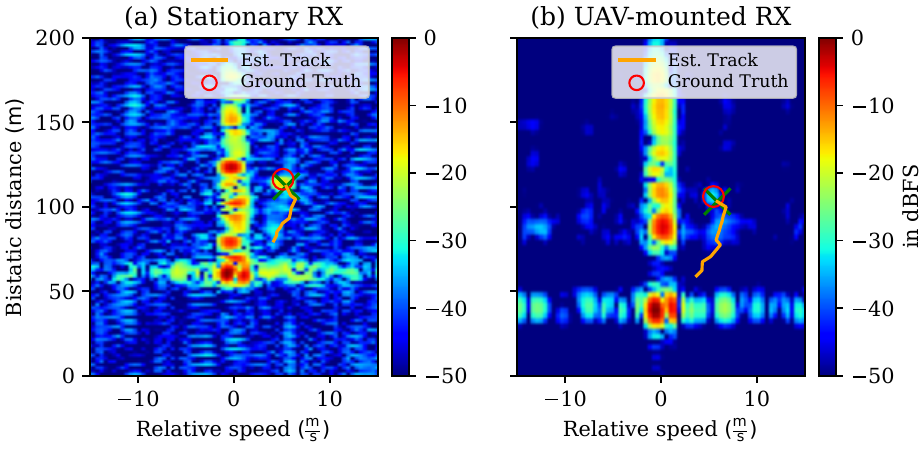}
\caption{Exemplary delay-Doppler maps with estimated tracks of the moving target ({\color{Purple}$\bullet$}) in a static environment for (a) a rooftop-mounted stationary sensor node ({\color{BurntOrange} $\bullet$}) and (b) a mobile sensor node at a hovering hexacopter ({\color{ForestGreen} $\bullet$}) in the measurement scenario shown in Fig.~\ref{measurement_site}. 
The red circle represents the ground-truth and the orange line is the estimated track of the VTOL.}
\label{measurement_delaydoppler}
\end{figure}
By applying well-known radar processing techniques, for instance as introduced in~\cite{10107507}, the \gls{vtol} aircraft can be detected and tracked along its flight track using bi-static delay and Doppler-shifts at individual receivers, as shown in Fig.~\ref{measurement_delaydoppler}.
This method includes following steps:
\begin{enumerate}
    \item point-wise division to obtain the complex channel transfer function using B2B-calibration data, 
    \item applying a fractional delay shift for drift compensation using high-resolution parameter estimation to compare the delay of the observed LoS delay and expected LoS delay calculated from the ground-truth antenna positions, 
    \item exponential background subtraction in the delay-Doppler domain to remove static clutter,
    \item target detection using a SO-CFAR detection algorithm,
    \item delay and Doppler parameter estimation to obtain off-grid estimations using parabolic interpolation, and 
    \item stabilizing the target tracking algorithm using Kalman filters to compensate short observation losses in prediction.
\end{enumerate}

\section{Conclusions}
This paper introduced a synchronized distributed channel sounding testbed designed for investigating and developing algorithms of multi-sensor ISAC networks. The testbed features both airborne and ground-based sensor nodes, with focus on modularity, flexibility, and highly accurate positioning of all participating nodes. The system's performance was validated through a real-world measurement campaign with a publicly available data set, demonstrating its capability to detect radar targets such as VTOL aircrafts in an A2A and A2G measurement scenarios using stationary ground-based nodes and mobile sensor nodes at an UAV swarm. 
Future work will include the public release of additional data sets, such as for V2X scenarios in parking lots, to support the development and evaluation of ISAC algorithms.
Moreover, we will explore advanced techniques to address the synchronization error compensation to ensure more robust and accurate results.

\bibliographystyle{IEEEtran}
\bibliography{paper}

\end{document}